\documentclass[twocolumn]{svjour3}
\smartqed
\usepackage{marvosym}
\usepackage{amsmath}
\usepackage{amssymb}
\usepackage{amsxtra}
\usepackage{amscd}
\usepackage{epsfig}
\usepackage{epstopdf}
\usepackage{float}
\usepackage{graphics}
\usepackage{graphicx,color}
\usepackage{hyperref}
\usepackage{latexsym}
\usepackage{longtable}
\usepackage{lscape}
\usepackage{mathptmx}
\usepackage{mathcomp}
\usepackage{multirow}
\usepackage{url}

\begin{document}
\title{Electron structure of superheavy elements Uut, Fl and Uup ($Z$=113 to 115).}
\author{V.~A.~Dzuba \and V.~V.~Flambaum}

\institute{ V.~A.~Dzuba (\Letter) \at 
	      School of Physics, University of New South Wales, Sydney 2052, Australia \\
            \email{V.Dzuba@unsw.edu.au}
            \and
            V.~V.~Flambaum \at
	      School of Physics, University of New South Wales, Sydney 2052, Australia \\
            \and Helmholtz Institute Mainz, Johannes Gutenberg University, , 55099 Mainz, Germany
}
\date{\today}
\maketitle
\begin{abstract}
We use recently developed method of accurate atomic calculations which combines linearized 
single-double coupled cluster method with the configuration interaction technique to calculate
ionisation potentials, excitation energies, static polarizabilities and valence electron densities 
for superheavy elements Uut, Fl and Uup ($Z$=113 to 115) and their ions. 
The accuracy of the calculations is controlled by comparing similar calculations for lighter 
analogs of the superheavy elements, Tl, Pb and Bi with experiment. The role of relativistic
effects and correlations is discussed and comparison with earlier calculations is presented.
\end{abstract}
% \pacs{31.15.vj,31.30.jg,11.30.Er}
\section{Introduction}

Study of superheavy elements (SHE, nuclear charge $Z >103$) is an important area of research motivated 
by the search of hypothetical stability island (see, e.g.~\cite{hofmann2010,Oganessian2013,turler2013,Fritz2013,schadel2015}). 
Experimental work focuses mostly on production and detection of SHE and study of their chemical
properties, while there is also some progress in the measurements of the atomic properties.
The heaviest element for which ionization potential has been measured is lawrencium 
($Z=103$)~\cite{sato2015measurement}.
There is good progress in the measurements of the frequencies of strong electric dipole transitions including
hyperfine structure and isotope shift for nobelium ($Z=102$)~\cite{Block2015,Sato2015}. 
For heavier atoms the information about their atomic properties comes mostly from atomic calculations. 
The calculations may help in experimental progress and predict chemical properties of SHE. 
Apart from that the SHE represent an interesting objects from pure theoretical point of view due to strong 
interplay between correlation and relativistic effects. For example, strong relativistic effects often lead to 
breaking similarity in chemical properties between SHE and their lighter analogs 
(see, e.g.~\cite{eliav2015,pershina2015,DinDzuFla08}).

In this paper we study SHE Uut, Fl and Uup ($Z$=113, 114 and 115). These elements are in the vicinity of
the hypothetical stability island and have relatively simple electron structure. They can be considered as 
atoms with three (Uut, Uup) or four (Fl) valence electrons. For such systems accurate calculations are
possible. The atoms were studied before mostly by multi-configuration Hartree-Fock and
and coupled cluster methods~\cite{E-E113,P-E113,E114a,E114b,E114c,E114d,E114e,E115}. 
In this paper we apply a recently developed method which
combines linearized single-double couple cluster (SD) and configuration interaction (CI) methods. 
We present  detailed study of the energy levels,  ionization potentials, static scalar polarizabilities, 
and electron densities for the superheavy elements and their ions. We discuss the role of relativistic
and correlation effects.

\section{Method of Calculation}

Atomic calculations always start from calculating the energy and the wave function of the ground state.
Then they may proceed to calculation of excited states, their lifetimes, transition amplitudes, etc.
Since there are no experimental data on atomic properties of superheavy elements considered in this paper
we should decide which properties should be calculated with higher priorities. Ionization potential (IP)
and polarizability of an atom are obviously important parameters since they determine its interaction 
with environment and thus, its chemical properties. This can be illustrated by the following example.
A way of detecting SHE and studying their chemical properties (e.g, volatility) is by observing the SHE absorption to an
inert surface. The energy of the interaction of the SHE with the surface is given by the formula~\cite{Pershina2001}
\begin{equation}
E(x) = - \frac{3}{16} \left(\frac{\epsilon -1}{\epsilon+2}\right)\frac{\alpha_{\rm at}}
{\left(\frac{1}{IP_{\rm slab}}+\frac{1}{IP_{\rm at}}\right)x^3},
\label{e:Ex}
\end{equation}
where $\alpha_{\rm at}$ is atomic polarizability and $IP_{\rm at}$ is its ionization potential; 
$\epsilon$ is the dielectric constant of the substance of the slab, $IP_{\rm slab}$ is the ionization 
potential of the slab atom, and $x$ is the surface-atom distance.

Other important characteristics of SHE include frequencies of strong electric dipole transitions from
the ground state since they are likely to be measured first. It is also instructive to have other excited states 
and electron densities and compare them to lighter analogs of SHE to study the role of relativistic effects.

Obtaining reliable results for superheavy elements (SHE) require accurate treatment of relativistic and 
many-body effects. The inclusion of the most important relativistic effects associated with the Dirac 
equation  is pretty straightforward. These relativistic effects are responsible for the difference between
SHE and lighter atoms. Inclusion of other relativistic corrections such as Breit interaction, quantum
electrodynamic (QED) corrections are also important for accurate results but they bring little difference
in electron structure of SHE compared to lighter atoms. The main challenge for the calculations and largest
uncertainty in the results comes from the treatment of the correlations. Accurate results can be 
obtained for atoms with simple electron structure with few valence electrons.  The highest accuracy
can be achieved for Fr-like and Ra-like atoms and ions which have only one or two valence electrons 
well separated from the core in space and on energy scale~\cite{DinEtAl08,DDFG08,DDF09,Dzu13,GD15}. In this paper
we consider atoms and ions with three and four valence electrons from 13th to 15th groups of the
periodic table. The method of treating correlations is presented in next section. Note that while
accurate treatment of correlations is very important for accurate results, the correlations for SHE are 
very similar to those of the lighter atoms and usually do not cause any significant difference in the atomic
properties of SHE.

\subsection{Treating the correlations. The SD+CI method.}

For accurate treatment of correlations we use the combinations of the single-double (SD)
coupled cluster method and the configuration interaction (CI) technique developed in
Ref.~\cite{Dzu-CI-SD14} (the SD+CI method). Here we give its brief summary.

All atomic electrons are divided into two categories, the core electrons, which occupy the closed-shell
core of the atom, and valence electrons, which occupy outermost open shells. The wave function of 
the valence electrons is assumed to be in the form
\begin{equation}
\Psi(r_1,\dots,r_{N_v}) = \sum_i c_i \Phi_i(r_1,\dots,r_{N_v}),
\label{e:Psi}
\end{equation}
where $N_v$ is the number of valence electrons ($N_v$=3 for Tl, E113,
Bi and E115, $N_v=4$ for Pb and E114), $\Phi_i(r_1,\dots,r_{N_v})$ is a single-determinant basis
state constructed from single-electron orbitals $\phi_j$. For example, for two valence electrons
\begin{equation}
\Phi_i(r_1,r_2) = \frac{1}{\sqrt{2}}\left(\phi_j(r_1)\phi_m(r_2)-\phi_m(r_1)\phi_j(r_2)\right).
\label{e:Phi2}
\end{equation}
The expansion coefficients $c_i$ in (\ref{e:Psi}) are found by solving the eigenvalue problem for the 
effective CI Hamiltonian,
\begin{equation}
\hat H^{\rm CI} = \sum_i^{N_v} \hat h_1(r_i) + \sum_{i<j}^{N_v} \hat
h_2(r_i,r_j),
\label{eq:CI}
\end{equation}
$\hat h_1$ and $\hat h_2$ are one and 
two-electron parts of the Hamiltonian. The one electron part is given by
\begin{equation}
\hat h_1 = c {\mathbf \alpha} {\mathbf p} + (\beta -1 )mc^2 +
V_{\rm core} + \hat \Sigma_1,
\label{eq:h1}
\end{equation}
where ${\mathbf \alpha}$ and $\beta$ are Dirac matrixes, $V_{\rm core}$ is
the self-consistent potential of the atomic core (including nuclear
part), $\hat \Sigma_1$ is the single-electron correlation operator
responsible for the correlation interaction of a valence electron with
the core. 

Two-electron part of the Hamiltonian is
\begin{equation}
\hat h_2(r_1,r_2) = \frac{e^2}{|r_1-r_2|} + \hat \Sigma_2(r_1,r_2),
\label{eq:h2}
\end{equation}
where $\hat \Sigma_2$ is the correlation correction to the Coulomb interaction between valence
electrons caused by core electrons.

The $\hat \Sigma_1$ and $\hat \Sigma_2$ operators represent core-valence correlations.
They can be calculated in the lowest, second order of the many-body perturbation theory (MBPT).
Inclusion of the second-order $\hat \Sigma$ into the CI calculations leads  to significant improvement 
of the accuracy of the calculations compared to the CI method with no $\hat \Sigma$. 
Further improvements in accuracy is achieved when
$\hat \Sigma_1$ and $\hat \Sigma_2$ are calculated to all orders using the SD 
approach~\cite{Dzu-CI-SD14,SD+CI}. This approach also includes core-core correlations.
The SD equations for the core do not depend on valence electrons and remain the same regardless 
of whether the SD and CI methods combined together or not. These equations can be found elsewhere 
(see, e.g.~\cite{Dzu-CI-SD14,SD+CI}). The SD equations for valence states in the SD+CI method 
can be written in the form 
\begin{eqnarray}
&&(\epsilon_0+\epsilon_b-\epsilon_m-\epsilon_n)\rho_{mnvb}  - \nonumber \\ 
&&\sum_{rs}g_{mnrs}\rho_{rsvb} - \sum_{rc}\tilde g_{cnrb}\tilde \rho_{mrvc} =
\label{e:sdv} \\
&&g_{mnvb} + \sum_{cd}g_{cdvb}\rho_{mncd} -  \sum_c g_{cnvb}\rho_{mc} 
 +  \nonumber \\
&& \sum_r g_{nmrv}\rho_{rb} -\sum_c g_{cmbv}\rho_{nc} 
+\sum_{rc}\tilde g_{cmrv}\tilde \rho_{nrbc}.   \nonumber 
\end{eqnarray}

Here parameters $g$ are Coulomb integrals 
\[ g_{mnab} = \int \int \psi_m^\dagger(r_1) \psi_n^\dagger(r_2)\frac{e^2}{r_{12}}
\psi_a(r_1)\psi_b(r_2)d\mathbf{r}_1d\mathbf{r}_2, \] 
parameters $\epsilon$ are the single-electron Hartree-Fock
energies, $\epsilon_0$ is an external parameter related to the energy of valence state
of interest~\cite{Dzu-CI-SD14}.
Coefficients $\rho_{ij}$ and $\rho_{ijkl}$ are the expansion
coefficients found by solving the SD equations. 
The tilde above $g$ or $\rho$ means the sum of direct and
exchange terms, e.g. 
\[ \tilde \rho_{nrbc} = \rho_{nrbc} - \rho_{nrcb}. \]
Indexes $a,b,c$ numerate states in atomic core, indexes
$m,n,r,s$ numerate states above the core, indexes $i,j,k,l$ numerate
any states.

Equations (\ref{e:sdv}) are solved iteratively starting from
\begin{eqnarray}
&& \rho_{mnij} =
\frac{g_{mnij}}{\epsilon_i+\epsilon_j-\epsilon_m-\epsilon_n}. \label{initial}
\end{eqnarray}
The right-hand side of  (\ref{e:sdv}) contains only terms which do not change in the iterations.
Corresponding $\rho_{ij}$ and $\rho_{ijkl}$ coefficients are found from solving the SD equations 
for the core (see Ref.~\cite{Dzu-CI-SD14,SD+CI} for more details).

After solving the SD equations, matrix elements for the $\hat \Sigma_1$ and $\hat \Sigma_2$
operators can be found by
\begin{eqnarray}
&&\langle m|\hat \Sigma_1| v \rangle  = \sum_{bn}\tilde g_{mban}\rho_{nb} + 
 \nonumber \\
&&\sum_{bnr}g_{mbnr}\tilde\rho_{nrvb}- \sum_{bcn}g_{bcvn}\tilde\rho_{mnbc},
\label{e:S1me}
\end{eqnarray}
and
\begin{eqnarray}
&&\langle mn |\hat \Sigma_2| vw \rangle =  \nonumber \\
&&\sum_{cd}g_{cdvw}\rho_{mncd} -  \sum_c \left( g_{cnvw}\rho_{mc} 
+ g_{cmwv}\rho_{nc} \right) + \nonumber \\
&&\sum_{rc} \left( g_{cnrw}\tilde \rho_{mrvc} 
+ g_{cmrv}\tilde \rho_{nrwc}  + g_{cnwr} \rho_{mrvc}
\right. \label{qscreen} \\   
&&+ g_{cmvr} \rho_{nrwc} - g_{cmwr} \rho_{nrcv} - \left. g_{cnvr}
  \rho_{mrcw} \right).  \nonumber  
\end{eqnarray}

A complete set of single-electron states is needed for solving the SD
equations (\ref{e:sdv}) and for construction of the
many-electron valence states (\ref{e:Psi}). We use the same B-spline
technique~\cite{B-splines} in both cases. Forty B-spline states
of the order of nine are calculated in a box of radius 40~$a_B$ in
the $V^{N-M}$ potential~\cite{Dzu05}. Here $N$ is total number of electrons and
$M$ is number of valence electrons. We include partial waves up to $l_{\rm max}=6$. 
Thirty lowest states in each partial wave are used for solving the SD equations for
the core and for valence states. The SD
equation for valence states are solved for few (usually three) states
above the core in each partial wave up to $l_{\rm max}=2$. The
second-order correlation potential $\hat \Sigma$ is used for higher
states. Fourteen states above the core in each partial wave up to
$l_{\rm max}=4$ are used in the CI calculations. With this choice of
the parameters the basis is sufficiently saturated.

\subsection{Study of relativistic effects}

Dominating relativistic effects $\left( \sim (Z\alpha)^2 \right)$ are included by solving Dirac-like equation
for since-electron orbitals. Breit and quantum electrodynamics (QED) corrections are also
included (see below). Choosing single-electron orbitals in the form
\begin{eqnarray}
\psi({\mathbf r}) = \frac{1}{r} \left( \begin{array}{c} f(r) \Omega_{\kappa,m}(\mathbf n) \\ 
 i\alpha g(r) \Omega_{-\kappa,m}(\mathbf n) \end{array} \right)
\label{e:psi}
\end{eqnarray}
leads to the following form of the Dirac equation (we use atomic units)
\begin{eqnarray}
&&\frac{\partial f}{\partial r} + \frac{\kappa}{r} f - \left[ 2 + \alpha^2(\epsilon - \hat V)\right]g = 0, \nonumber \\
&&\frac{\partial g}{\partial r} - \frac{\kappa}{r} g + (\epsilon - \hat V)f = 0.
\label{e:Dirac}
\end{eqnarray}
Here $\mathbf{n} \equiv \mathbf{r}/r$, $\kappa =(-1)^{l+j+1/2}(j+1/2)$ defines electron orbital 
angular momentum $l$ and total angular momentum $j$, $\alpha$ is the fine structure constant,
and $\hat V$ is the potential. Potential includes nuclear and electron parts. Electron part
includes direct and exchange terms.  Fine structure constant $\alpha$ serves as a measure of 
relativistic effect. Non-relativistic limit corresponds to $\alpha=0$. Varying the value of $\alpha$ 
in (\ref{e:Dirac}) and corresponding computer codes allows us to study the role of relativistic
effects in atoms. For example, it was used to search for the manifestations of the hypothetical 
space-time variation of the fine structure constant in atomic spectra (see, e.g. 
\cite{DzuFlaWeb99,DzuFlaWeb99a,DzuFlaCJP}).

We include Breit interaction in the zero momentum transfer approximation. Corresponding 
Hamiltonian is
\begin{equation}
\hat H^{B}=-\frac{\mbox{\boldmath$\alpha$}_{1}\cdot \mbox{\boldmath$\alpha$}_{2}+
(\mbox{\boldmath$\alpha$}_{1}\cdot {\bf n})
(\mbox{\boldmath$\alpha$}_{2}\cdot {\bf n})}{2r} \ .
\label{e:Breit}
\end{equation}  
Here ${\bf r}={\bf n}r$, $r$ is the distance between electrons, and 
$\mbox{\boldmath$\alpha$}$ is the Dirac matrix. The Hamiltonian (\ref{e:Breit})
includes magnetic interaction and retardation. 
Similar to Coulomb interaction, we determine the self-consistent 
Hartree-Fock contribution arising from Breit interaction by adding Breit term to the
self-consistent Hartree-Fok potential, 
\begin{equation}
\hat V=V^{C}+V^{B} \ .
\label{e:Vb}
\end{equation}  
Here $V^{C}$ is the Coulomb potential, $V^B$ is the Breit potential.
Iterating Hartree-Fock equations with the Breit term in the potential corresponds to
the inclusion of the important relaxation effect (see, e.g.~\cite{QED-relaxation}). Note
that this also leads to inclusion of the higher-order Breit corrections which have no
physical meaning. In principle, they can be removed by simple rescaling 
procedure~\cite{DzuEtAl01a,DzuFlaSaf06}. However, in practice they are small 
and do not affect the results.

Quantum electrodynamics radiative corrections to the energies (Lamb shifts) 
are accounted for by the use of the 
radiative potential introduced in Ref. \cite{radpot}. This potential has the form
\begin{equation}
V_{\rm rad}(r)=V_U(r)+V_g(r)+ V_e(r) \ ,
\end{equation}
where $V_U$ is the Uehling potential, $V_g$ is the potential arising from the 
magnetic formfactor, and $V_e$ is the potential arising from the electric formfactor.
As for the case of the Breit interaction, the relaxation effect is important and included 
by iterating the Hartre-Fock equations with the radiative potential added to the Hartree-Fock 
potential.
%------------------------------------------------------------------------------------------------------------
\begin{table}
\begin{center}
\caption{\label{t:Hg-Bi}
Calculated and experimental ionization potentials (IP, in cm$^{-1}$) and
excitation energies (EE, in cm$^{-1}$) for the ground and 
lowest excited states of Tl, Pb and Bi atoms. 
Other theoretical data are taken from Ref.~\cite{E-E113} for Tl, Ref.~\cite{E114d} for Pb and Ref.~\cite{E115} for Bi.
Experimental values are taken from the NIST database~\cite{NIST}.} 
%\begin{ruledtabular}
\begin{tabular}{llll rrr}
\multicolumn{1}{c}{Atom}&&
\multicolumn{2}{c}{State}&
\multicolumn{2}{c}{$E_{\rm calc}$}&
\multicolumn{1}{c}{$E_{\rm expt}$} \\
&&&&
\multicolumn{1}{c}{this work}&
\multicolumn{1}{c}{other} & \\
\hline\noalign{\smallskip}

Tl & IP  & $6s^{2}6p$   & $^2$P$^o_{1/2}$ & 49228 & 48575 & 49266 \\
   & EE  & $6s^{2}6p$   & $^2$P$^o_{3/2}$ &  7897 & 7710 &  7793 \\
   &     & $6s^{2}7s$   & $^2$S$_{1/2}$   & 26713 & 26456 & 26478 \\
   &     & $6s^{2}8p$   & $^2$P$^o_{1/2}$ & 34486 & & 34160 \\
   &     & $6s^{2}8p$   & $^2$P$^o_{3/2}$ & 35456 & & 35161 \\
   &     & $6s^{2}6d$   & $^2$D$_{ 3/2}$  & 36348 & & 36118 \\
   &     & $6s^{2}6d$   & $^2$D$_{ 5/2}$  & 36466 & & 36200 \\
   &     & $6s^{2}8s$   & $^2$S$_{1/2}$   & 40932 & & 38746 \\
   &     & $6s^{2}5f$   & $^2$F$^o_{5/2}$ & 41982 & & 42318 \\
   &     & $6s^{2}5f$   & $^2$F$^o_{7/2}$ & 42011 & & 42318 \\

Pb & IP & $6s^{2}6p^{2}$  & $^1$S$_0$ &  58980 & 59276 & 59819 \\
   & EE & $6s^{2}6p^{2}$  & $^3$P$_1$ &   7538 &  7531 & 7819 \\
   &    & $6s^{2}6p^{2}$  & $^3$D$_2$ &  10527 &  10307 & 10650 \\
   &    & $6s^{2}6p^{2}$  & $^3$D$_2$ &  21033 &  20853 & 21457 \\
   &    & $6s^{2}6p^{2}$  & $^1$S$_0$ &  29438 &  29259 & 29466 \\
   &    & $6s^{2}7s6p$   & $^1$S$^o_0$ & 35737 &  34405 & 34960 \\
   &    & $6s^{2}7s6p$   & $^3$P$^o_1$ & 36099 &  34711 &35287 \\

Bi & IP & $6p^{3}$  & $^4$P$^o_{ 3/2}$ &    58502 & 58656 &  58762 \\
   & EE & $6p^{3}$  & $^4$D$^o_{ 3/2}$ &    11673 & &  11419 \\
   &    & $6p^{3}$  & $^2$D$^o_{ 5/2}$ &    15750 & &  15438 \\
   &    & $6p^{3}$  & $^2$P$^o_{ 1/2}$ &    21710 & &  21661 \\
   &    & $7s6p^{2}$& $^2$S$_{ 1/2}$ &      32542 & &  32588 \\
   &    & $6p^{3}$  & $^4$D$^o_{ 3/2}$ &    33486 & &  33164 \\
   &    & $6p^{2}7p$& $^2$P$^o_{ 1/2}$ &    42276 & &  41125 \\
   &    & $6p^{2}8p$& $^2$P$^o_{ 3/2}$ &    43971 & &  42940 \\
   &    & $6p^{2}6d$& $^2$D$_{ 3/2}$   &    45086 & &  43912 \\
   &    & $6p^{2}6d$& $^2$D$_{ 5/2}$   &    45926 & &  44816 \\
   &    & $6p^{2}7s$& $^4$P$_{ 3/2}$   &    44411 & &  44865 \\
   &    & $6p^{2}7s$& $^2$P$_{ 1/2}$   &    45587 & &  45915 \\
\hline\noalign{\smallskip}
\end{tabular}
%\end{ruledtabular}
\end{center}
\end{table}
%------------------------------------------------------------------------------------------------------------
Table \ref{t:Hg-Bi} compares calculated ionisation potentials and lowest excitation 
energies of Tl, Pb and Bi atoms with experimental data. Correlations and relativistic
effects are included in all calculations as has been described above. 
The difference between theory and experiment is on the level of $\sim 1 $\%.

WE also present in Table~\ref{t:Hg-Bi} the results of coupled-cluster (CC) calculations of the energy
levels of Tl, Pb and Bi taken from the works devoted to superheavy elements E113~\cite{E-E113},
E114~\cite{E114d} and E115~\cite{E115}. All calculations are done with the same method and probably 
with the same set of computer codes. The most detailed data are presented for Pb. Comparison 
shows that both calculations have very similar deviations from the experimental data. On the other
hand, the difference between two calculations is significantly smaller that the difference between
theory and experiment at least for the lowest states. This is important observation since the calculations
are done with the use of very different techniques. Good agreement between different
calculations adds to the reliability of the results.

\subsection{Calculation of polarizabilities}

\begin{table*}
\begin{center}
\caption{\label{t:pol}
Contributions to the static scalar  polarizabilities $\alpha_0$ ($\alpha_0=\alpha_c+\alpha_{cv}+\alpha_v$)
of the ground state of Tl, Pb and Bi atoms (in $a_B^3$) and comparison with earlier calculations or
measurements. See formula (\ref{e:alpha-cv}) and discussion below it for definitions 
of $\alpha_c$, $\alpha_{cv}$, and $\alpha_v$. }
\begin{tabular}{lll rrrr l}
&&&\multicolumn{4}{c}{Present work}&\multicolumn{1}{c}{Other} \\
\multicolumn{1}{c}{Atom}&
\multicolumn{2}{c}{State}&
\multicolumn{1}{c}{$\alpha_c$}&
\multicolumn{1}{c}{$\alpha_{cv}$}&
\multicolumn{1}{c}{$\alpha_v$}&
\multicolumn{1}{c}{$\alpha_0$}&
\multicolumn{1}{c}{$\alpha_0$} \\
\hline\noalign{\smallskip}

Tl  &  $6s^{2}6p$        & $^2$P$^o_{1/2}$ &  4.98 & -0.11 & 42.91 & 47.78 &   
 49.2\cite{KPJ01}, 51.6\cite{Fleig}, 51.3(5.4)\cite{Guella}   \\ 
 
Pb &  $6s^{2}6p^{2}$  & $^1$S$_0$      &  3.63 & -0.44 & 40.85 & 44.04 &  
 47.3\cite{Thierfelder},47.1(7.0)\cite{Thierfelder},  46(11)\cite{Doolen}\\ 
 
Bi  &  $6p^3$              & $^4$P$^o_{3/2}$ & 11.22 & -5.43 & 38.83 & 44.62 &  
 50(12)\cite{Doolen} \\ 
\hline\noalign{\smallskip}
\end{tabular}
%\footnotetext[1]{Experiment}
\end{center}
\end{table*}

Polarisability is an important characteristic of an atom defining its interaction with external field. 
E.g., second-order Stark shift is given by $\delta E = -\frac{1}{2}\alpha\varepsilon^2$, where $E$ is atomic
energy, $\alpha$ is polarisability and $\varepsilon$ is electric field. For atoms with total angular momentum
$J \ge 1$ polarisability is given by the sum of two terms
\begin{equation}
\alpha = \alpha_0 + \alpha_2 \frac{3M^2 - J(J+1)}{J(2J-1)},
\label{e:alpha02}
\end{equation}
where $\alpha_0$ is scalar polarisability,  $\alpha_2$ is tensor polarisability and $M$ is projection
of $J$.

When many-electron atom is treated by a configuration interaction technique all atomic electrons are 
divided into two groups. Most go to the closed-shell core, and few remaining are treated as valence 
electrons. As a consequence, there are three contributions to the scalar polarisability, core contribution, 
core-valence contribution, and valence contribution
\begin{equation}
	\alpha_0 = \alpha_c + \alpha_{cv} + \alpha_v.
\label{e:alpha-cv}
\end{equation}
Core-valence contributions arise due to the interference between core and valence terms. 
Calculation of the core polarisability is affected by the presence of the valence electrons due to the 
Pauli principle. States, occupied by valence electrons must be excluded from the summation over 
complete set of states.

Another consequence of dividing electrons into core and valence categories  is the effect of core 
polarisation by external field. This effect is usually included within the so-called random-phase 
approximation (RPA). The RPA equations for single-electron orbitals $\phi$ can be written as 
\begin{equation}
(\hat H^{\rm HF} - \epsilon)\delta \phi = -(\hat F + \delta V)\phi,
\label{e:RPA}
\end{equation}
where $\hat H^{\rm HF}$ is the Hartree-Fock Hamiltonian, $\delta \phi$ is the correction to the 
single-electron orbital $\phi$ due to external field, $\hat F$ is the operator of the external field,
and $\delta V$ is the correction to the self-consistent Hartree-Fock potential due to the external 
field. In case of calculation of the static polarizabilities, the external field is static electric field and 
$\hat F$ is the electric dipole operator, $\hat F \equiv \hat d = -e\hat r$.

Core contribution to the polarisability is given by
\begin{equation}
\alpha_c = \frac{2}{3} \sum_{an} \frac{\langle a||\hat d ||n\rangle\langle a || \hat d + \delta V || n \rangle}
{\epsilon_n - \epsilon_a}
\label{e:alphac}
\end{equation}
Summation goes over core states $a$ and complete set of single-electron states above the core $n$.
Note that the RPA correction is included into only one of two matrix elements 
(see, e.g. Ref.~\cite{pol-rev} for more detailed discussion).

Expression (\ref{e:alphac}) does not take into account the presence of valence electrons. States occupied
by valence electrons must be excluded from the summation over excited states $n$ in (\ref{e:alphac}).
This is not a trivial task since states $n$ are just single-electron basis states but not real physical states. 
The solution within the CI approach is the following.   Recall that the many-electron wave function for
valence electrons has the form (\ref{e:Psi}).
Then fractional occupation number $w_n$ for a single-electron orbital $\phi_n$ can be found as a sum
\begin{equation}
w_n = \sum_i c_{in}^2,
\label{e:w}
\end{equation}
where additional index $n$ for expansion coefficients means  that summation goes only over 
those states $\Phi_i$ in (\ref{e:Psi}) which contain 
single-electron orbital $\phi_n$ in the definition of $\Phi_i$ (e.g., formula (\ref{e:Phi2}) for the 
two-valence-electrons case). Final occupation number for the orbital $\phi_n$ is the ratio of $w_n$ to the 
maximum possible number of electrons in state $n$ which is $2j_n+1$. In the end, the sum of the core 
and core-valence contributions to atomic polarisability can be written as
\begin{eqnarray}
&&\alpha_c +\alpha_{vc} = \label{e:alphacv} \\
&&\frac{2}{3} \sum_{an} \frac{\langle a||\hat d ||n\rangle\langle n || \hat d + 
\delta V || a \rangle}{\epsilon_n - \epsilon_a}\left(1 - \frac{w_n}{2j_n+1}\right).
\nonumber
\end{eqnarray}
Valence contribution to the polarisability is given by
\begin{equation}
\alpha_v = \frac{2}{3(2J_0+1)} \sum_{N} \frac{|\langle 0 || \hat D^{\rm RPA} || N \rangle|^2}
{E_N - E_0}.
\label{e:alphav}
\end{equation}
The summation goes over complete set of many-electron valence states $N$ which are the eigenstates 
of the CI hamiltonian and have a form of (\ref{e:Psi}). $E_N$ is the energy of state $N$, $E_0$ and $J_0$ 
are the energy and total angular momentum of the state $|0\rangle$, $\hat D^{\rm RPA}$ is the many-electron operator
of electric dipole moment, $\hat D^{\rm RPA} = \sum_i^{N_v}(\hat d_i+\delta V_i)$.

Performing summation over many-electron states in (\ref{e:alphav}) is a challenging task since most
if not all of the eigenstates of the CI Hamiltonian need to be included for accurate results. We use
the technique introduced by Dalgarno and Lewis ~\cite{DalLew55} to perform the summation. 
In this approach a correction to the ground state many electron wave function is introduced in a form
\begin{equation}
|\tilde 0 \rangle = \sum_N |N\rangle\frac{\langle N||\hat D^{\rm RPA}||0\rangle}{E_N-E_0}.
\label{e:delta0}
\end{equation}
Since all states $N$ and $0$ in (\ref{e:delta0}) are the eigenstates of the CI Hamiltonian, the correction
(\ref{e:delta0}) satisfies the equation
\begin{equation}
(\hat H^{\rm CI} - E_0)|\tilde 0\rangle = -D^{\rm RPA}|0\rangle.
\label{e:delta0e}
\end{equation}
In the CI technique the correction $|\tilde 0 \rangle$ has a form of (\ref{e:Psi}) and equation (\ref{e:delta0e})
is reduced to the system of linear equations for expansion coefficients $c_i$. After solving the equations we
calculate valence part of the polarisability using the formula
\begin{equation}
\alpha_v =\frac{2}{3(2J_0+1)}\langle 0 ||\hat D^{\rm RPA}||\tilde 0\rangle.
\label{e:alphaDL}
\end{equation} 
Calculation of tensor polarisability $\alpha_2$ (see Eq.~(\ref{e:alpha02}) is similar. There is no core 
contribution because total angular momentum of the closed shell core is zero. Calculation of the valence
contribution involves performing the same summation as in (\ref{e:alphav}). It can be done ones for both
polarizabilities, $\alpha_0$ and $\alpha_2$. The formula for $\alpha_2$ differs from (\ref{e:alphav}) by
angular coefficients only and can be found elsewhere (see, e.g.~\cite{pol-rev}).

Static scalar polarizabiliries of Tl, Pb and Bi are presented in Table~\ref{t:pol}. Bi atom has also 
tensor polarisability of the ground state which is found to be $\alpha_2 = -5.38 \ a_B^3$.
As can be seen from the data all terms, core, core-valence 
and valence, give significant contribution to the scalar polarizability and there is good agreement 
between present calculations and available experimental data and earlier calculations. Judging
by the differences in results the accuracy of present calculations is within few percent.

\subsection{Density of valence electrons}

In the configuration interaction calculations many-electron wave function for the valence electrons has a form
of (\ref{e:Psi}) in which single-determinant basis states $\Phi_i$ are constructed from single-electron
states of the form (\ref{e:psi}). Therefore, density of valence electrons as a function of distance can
be calculated as
\begin{equation}
\rho_v(r) = \sum_i c^2_i \sum_{n=1}^{N_v} \left( f_n^2(r)+\alpha^2 g^2_n(r)\right).
\label{e:rho}
\end{equation}
This expression is the most accurate one. It includes correlations and relativistic effects. To study the
effect of correlations one should compare (\ref{e:rho}) with the electron density calculated in the HF 
approximation using the $V^N$ potential (all atomic electrons are included into the self-consistent HF 
procedure),
\begin{equation}
\rho_v(r)_{\rm HF} =  \sum_{n=1}^{N_v} \left( f_n^2(r)+\alpha^2 g^2_n(r)\right),
%\rho_v(r)_{\rm HF} =  \sum_{n=1}^{N_v}  f_n^2(r)+\alpha^2 g^2_n(r),
\label{e:rhoHF}
\end{equation}
where $f_n$ and $g_n$ are single-electron HF wave functions of the valence electrons.
Non-relativistic limit for the density can be obtained if the calculations are conducted with
$\alpha=0$.

\section{Results and discussion}

\begin{table}
\begin{center}
\caption{\label{t:E113}
Calculated excitation energies ($E$, cm$^{-1}$), ionisation potentials
(I.P.) and $g$-factors for the lowest states of the E113 atom and
E113$^+$ and E113$^{++}$ ions.} 
%\begin{ruledtabular}
\begin{tabular}{ll rlr}
&&\multicolumn{2}{c}{This work}&
\multicolumn{1}{c}{Other}\\
\multicolumn{2}{c}{State}&
%\multicolumn{1}{c}{Term}&
\multicolumn{1}{c}{$E$}&
\multicolumn{1}{c}{$g$}&
\multicolumn{1}{c}{$E$~\cite{E-E113}}\\
\hline\noalign{\smallskip}
\multicolumn{4}{c}{E113, I.P. = 7.37 eV}\\

 $7s^{2}7p$              & $^2$P$^o_{ 1/2}$ &        0 &  0.6667 & 0 \\
 $7s^{2}7p$              & $^2$P$^o_{ 3/2}$ &    24758 &  1.3333 & 22528 \\
 $7s^{2}8s$              & $^2$S$_{ 1/2}$ &    36665 &  1.9997 & 35124 \\
 $7s^{2}8p$              & $^2$P$^o_{ 1/2}$ &    44346 &  0.6667 & \\
 $7s^{2}8p$              & $^2$P$^o_{ 3/2}$ &    47104 &  1.3333 & \\
 $7s^{2}7d$              & $^2$D$_{ 3/2}$ &    48768 &  0.8000 & \\
 $7s^{2}7d$              & $^2$D$_{ 5/2}$ &    49176 &  1.2000 & \\
 $7s^{2}9s$              & $^2$S$_{ 1/2}$ &    49774 &  2.0001 & \\
 $7s^{2}9p$              & $^2$P$^o_{ 1/2}$ &    51375 &  0.6667 & \\
 $7s^{2}6f$              & $^2$F$^o_{ 5/2}$ &    52225 &  0.8571 & \\
 $7s^{2}6f$              & $^2$F$^o_{ 7/2}$ &    52231 &  1.1429 & \\
 $7s^{2}9p$              & $^2$P$^o_{ 3/2}$ &    52485 &  1.3333 & \\

\multicolumn{4}{c}{E113$^{+}$, I.P. = 23.6 eV}\\

 $7s^{2}$                & $^1$S$_0$ &        0 &  0.0000 & \\
 $7s7p$                  & $^3$P$^o_0$ &    60042 &  0.0000 & 61245 \\
 $7s7p$                  & $^3$P$^o_1$ &    63187 &  1.3969 & 66868 \\
 $7s7p$                  & $^3$P$^o_2$ &    97312 &  1.4999 & 82208 \\
 $7s7p$                  & $^1$P$^o_1$ &   104560 &  1.1024 & 107182 \\
 
\multicolumn{4}{c}{E113$^{++}$, I.P. = 33.5 eV}\\

 $7s$                    & $^2$S$_{ 1/2}$ &        0 &  2.0000 & \\
 $7p$                    & $^2$P$^o_{ 1/2}$ &    70054 &  0.6667 & 74779 \\
 $7p$                    & $^2$P$^o_{ 3/2}$ &   117593 &  1.3333 & 111044 \\
\hline\noalign{\smallskip}
 
\end{tabular}
%\end{ruledtabular}
\end{center}
\end{table}

\begin{table}
\begin{center}
\caption{\label{t:E114}
Calculated excitation energies ($E$, cm$^{-1}$), ionisation potentials
(I.P.) and $g$-factors for lowest states of the Fl atom (E114) and its ion.} 
%\begin{ruledtabular}
\begin{tabular}{ll rlr}
&&\multicolumn{2}{c}{This work}&
\multicolumn{1}{c}{Other}\\
\multicolumn{2}{c}{State}&
%\multicolumn{1}{c}{Term}&
\multicolumn{1}{c}{$E$}&
\multicolumn{1}{c}{$g$}&
\multicolumn{1}{c}{$E$~\cite{E114d}}\\
\hline\noalign{\smallskip}
\multicolumn{4}{c}{E114, I.P. = 8.37 eV}\\

 $7s^{2}7p^{2}$          & $^1$S$_0$ &        0 &  0.0 & 0 \\
 $7s^{2}7p^{2}$          & $^3$P$_1$ &    27316 &  1.499 & 26342 \\
 $7s^{2}7p^{2}$          & $^3$D$_2$ &    29149 &  1.190 & 28983 \\
 $7s^{2}8s7p$            & $^1$S$^o_0$ & 44036 &  0.0 & 43111 \\
 $7s^{2}8s7p$            & $^3$P$^o_1$ & 44362 &  1.341 & 43441 \\
 $7s^{2}7p8p$            & $^3$D$_1$ &    51834 &  0.667 & 51302 \\
 $7s^{2}7p8p$            & $^1$S$_0$ &    53149 &  0.0       & 52487 \\
 $7s^{2}7p8p$            & $^3$P$_1$ &    55414 &  1.5       & 54647 \\
 $7s^{2}7p8p$            & $^3$D$_2$ &    55191 &  1.170 & 54814 \\
 $7s^{2}7p7d$            & $^3$F$^o_2$ & 56988 &  0.769 & \\
 $7s^{2}7p7d$            & $^1$P$^o_1$ & 57244 &  1.045 & \\
 $7s^{2}9s7p$            & $^1$S$^o_0$ & 57367 &  0.0   & \\
 $7s^{2}7p7d$            & $^3$D$^o_2$ & 57413 &  1.282 & \\
 $7s^{2}7p7d$            & $^3$F$^o_3$ & 57481 &  1.113 & \\
 $7s^{2}7p6f$             & $^5$G$_3$ &    60291 &  0.876 & \\
 $7s^{2}7p6f$             & $^3$F$_3$ &    60298 &  1.151 & \\
 $7s^{2}7p6f$             & $^3$G$_4$ &    60311 &  1.084 & \\

\multicolumn{4}{c}{E114$^{+}$, I.P. = 17.0 eV}\\

 $6d^{10}7s^{2}7p$              & $^2$P$^o_{ 1/2}$ &          0 &  0.667 & \\
 $6d^{10}7s^{2}7p$              & $^2$P$^o_{ 3/2}$ &  40380 &  1.333 & \\
 $6d^{10}7s7p^{2}$              & $^2$S$_{ 1/2}$ &      73746 &  2.250& \\
 $6d^{10}7s^{2}8s$              & $^2$S$_{ 1/2}$ &      74772 &  2.019 & \\

 $6d^97s^{2}7p^2$              & $^2$S$_{ 5/2}$ &      75133 &  1.211 & \\
$6d^97s^{2}7p^2$              & $^2$S$_{ 3/2}$ &      110250 &  0.876 & \\
\hline\noalign{\smallskip}

\end{tabular}
%\end{ruledtabular}
\end{center}
\end{table}

\begin{table}
\begin{center}
\caption{\label{t:E115}
Calculated excitation energies ($E$, cm$^{-1}$), ionisation potentials (I.P.), and $g$-factors for
the lowest states of E115, E115$^+$ and E115$^{++}$.} 
%\begin{ruledtabular}
\begin{tabular}{ll rl}
\multicolumn{2}{c}{State}&
%\multicolumn{1}{c}{Term}&
\multicolumn{1}{c}{$E$}&
\multicolumn{1}{c}{$g$}\\
\hline\noalign{\smallskip}
\multicolumn{4}{c}{E115, I.P. = 5.56 eV}\\
 $7p^{3}$                & $^2$P$^o_{ 3/2}$ &        0 &  1.4039 \\
 $7p^{2}8s$              & $^2$S$_{ 1/2}$ &    18377 &  2.0427 \\
 $7p^{2}8p$              & $^2$P$^o_{ 1/2}$ &    28382 &  0.6668 \\
 $7p^{2}8p$              & $^2$P$^o_{ 3/2}$ &    31716 &  1.3449 \\
 $7p^{2}7d$              & $^2$D$_{ 3/2}$ &    33092 &  0.7987 \\
 $7p^{2}7d$              & $^2$D$_{ 5/2}$ &    33273 &  1.2007 \\
 $7p^{2}9s$              & $^2$S$_{ 1/2}$ &    34332 &  2.0078 \\
 $7p^{3}$                & $^2$P$^o_{ 3/2}$ &    35484 &  1.4295 \\
 $7p^{2}9p$              & $^2$P$^o_{ 1/2}$ &    36884 &  0.6667 \\
 $7p^{2}9p$              & $^2$P$^o_{ 3/2}$ &    38127 &  1.3338 \\
\multicolumn{4}{c}{E115$^{+}$, I.P. = 18.4 eV}\\

 $7p^{2}$                & $^1$S$_0$ &        0 &  0.0000 \\
 $7p^{2}$                & $^3$P$_1$ &    42137 &  1.4999 \\
 $7p^{2}$                & $^3$D$_2$ &    45530 &  1.1931 \\
 $8s7p$                  & $^1$S$^o_0$ &    80575 &  0.0000 \\
 $8s7p$                  & $^3$P$^o_1$ &    81159 &  1.3412 \\
 $7p^{2}$                & $^3$D$_2$ &    93328 &  1.3066 \\
 $7p8p$                  & $^3$D$_1$ &    96781 &  0.6669 \\
 $7p7d$                  & $^3$F$^o_2$ &    99056 &  0.7731 \\
 $7p8p$                  & $^1$S$_0$ &    99066 &  0.0000 \\
 $7p7d$                  & $^1$P$^o_1$ &   100050 &  0.8036 \\
 $7p7d$                  & $^3$D$^o_2$ &   100282 &  1.2754 \\
 $7p7d$                  & $^3$F$^o_3$ &   100521 &  1.1167 \\
 $7p^{2}$                & $^1$S$_0$ &   102055 &  0.0000 \\
 $7p8p$                  & $^3$P$_1$ &   105263 &  1.4895 \\
 $7p8p$                  & $^3$D$_2$ &   105556 &  1.1691 \\

\multicolumn{4}{c}{E115$^{++}$, I.P. = 27.7 eV}\\

 $7p$    & $^2$P$^o_{ 1/2}$ &        0 &  0.6667 \\
 $7p$    & $^2$P$^o_{ 3/2}$ &    55538 &  1.3333 \\
 $8s$    & $^2$S$_{ 1/2}$  &   109339 &  2.0000 \\
 $7d$    & $^2$D$_{ 3/2}$  &   125854 &  0.8000 \\
 $7d$    & $^2$D$_{ 5/2}$ &   128850 &  1.2000 \\
 $8p$    & $^2$P$^o_{ 1/2}$ &   132480 &  0.6667 \\
 $8p$    & $^2$P$^o_{ 3/2}$ &   145164 &  1.3333 \\
 $6f$    & $^2$F$^o_{ 5/2}$ &   152111 &  0.8571 \\
 $6f$    & $^2$F$^o_{ 7/2}$ &   152167 &  1.1429 \\
\hline\noalign{\smallskip}

\end{tabular}
%\end{ruledtabular}
\end{center}
\end{table}

%---------------------------------------------------------------------------------------------------------
\begin{table}
\begin{center}
\caption{\label{t:IP}
Ionization potential of Tl, Pb, Bi and superheavy elements E113, E114, E115 including ions (eV). 
Comparison with experiment and other calculations.}
%\begin{ruledtabular}
\begin{tabular}{lccl}
\multicolumn{1}{c}{Atom/}&
\multicolumn{1}{c}{Expt.}&
\multicolumn{1}{c}{Present}&
\multicolumn{1}{c}{Other} \\
\multicolumn{1}{c}{Ion}&
\multicolumn{1}{c}{\cite{NIST}}&
\multicolumn{1}{c}{work}& \\
\hline\noalign{\smallskip}

Tl                   & 6.108 & 6.103 & 6.022~\cite{E-E113}; 6.096~\cite{P-E113} \\
Pb                  &7.416 & 7.312 & 7.035~\cite{E114a}; 7.35~\cite{E114b}; 7.349~\cite{E114d}; 7.484~\cite{E114d} \\
Bi                   &7.285 & 7.253 & 7.272~\cite{E115} \\
E113              && 7.37 & 7.306~\cite{P-E113}; 7.29~\cite{E-E113} \\
E113$^+$      && 23.6 & 23.9~\cite{E-E113} \\
E113$^{++}$  && 33.5 & 33.6~\cite{E-E113} \\
E114              && 8.37 &  8.28~\cite{E114a}; 8.54~\cite{E114b}; 8.539~\cite{E114d} \\
E114$^{+}$    && 17.0 &  17.2~\cite{E114a}; 16.871~\cite{E114d}\\
E115              && 5.56 &  5.574~\cite{E115} \\
E115$^{+}$    && 18.4 & 18.360~\cite{E115} \\
E115$^{++}$  && 27.7 &  \\
\hline\noalign{\smallskip}

\end{tabular}
%\tablenotetext[1]{Ref.~\cite{P-E113}}
%\end{ruledtabular}
\end{center}
\end{table}

%---------------------------------------------------------------------------------------------------------
\begin{table}
\begin{center}
\caption{\label{t:SHEpol}
Static scalar  polarizabilities $\alpha_0$ of SHE (in a.u.). Comparison with other calculations.}
%\begin{ruledtabular}
\begin{tabular}{lllll}
\multicolumn{1}{c}{Atom}&
\multicolumn{2}{c}{Ground}&
\multicolumn{1}{c}{Present}&
\multicolumn{1}{c}{Other} \\
&\multicolumn{2}{c}{state}&
\multicolumn{1}{c}{work}& \\
\hline\noalign{\smallskip}

%Cn &  $7s^{2}$           & $^1$S$_0$       &  26.0 & \\

E113  &  $7s^{2}7p$        & $^2$P$^o_{1/2}$ &  28.8 & 29.85~\cite{P-E113} \\
 
Fl &  $7s^{2}7p^{2}$  & $^1$S$_0$      &  31.4 & 31.0~\cite{E114c}; %30.72; 30.28; 
34.4~\cite{Nash2005}; 30.59~\cite{E114e} \\
 
E115  &  $7p^3$              & $^4$P$^o_{3/2}$ &  70.5 & \\
\hline\noalign{\smallskip}

\end{tabular}
%\end{ruledtabular}
\end{center}
\end{table}
%---------------------------------------------------------------------------------------------------------

\begin{figure}[th]
\begin{center}
\resizebox{0.42\textwidth}{!}{
\includegraphics{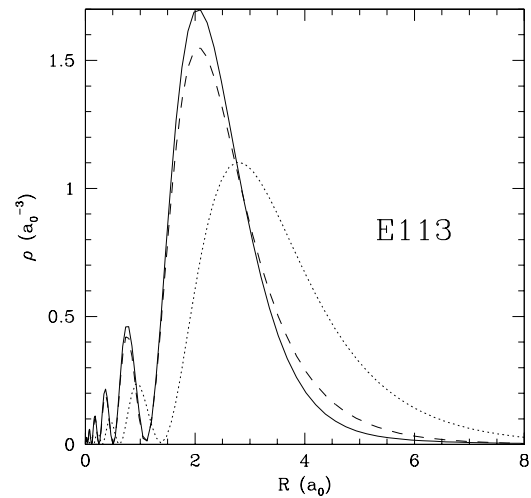}}
\end{center}
\caption{Density of valence electrons for the ground state ($7s^2 7p_{1/2} \ ^2$P$^o_{1/2}$) of Uut (E113).
Solid line - CI wave function (\ref{e:rho}). Dashed line - density in the relativistic Hartree-Fock approximation 
(\ref{e:rhoHF}). Dotted line is the Hartree-Fock density in the non-relativistic limit.}
\label{f:Uut}
\end{figure}

\begin{figure}
\resizebox{0.42\textwidth}{!}{
\includegraphics{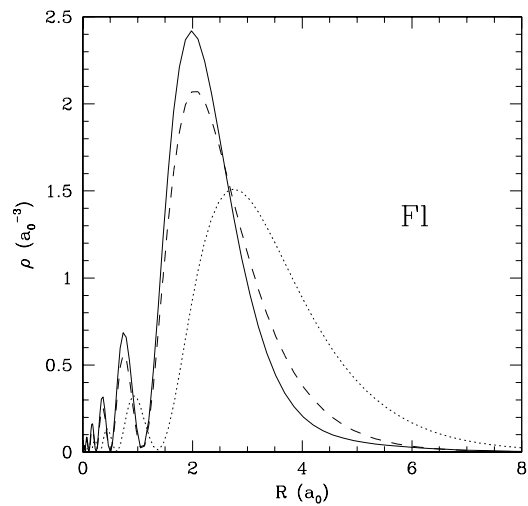}}
\caption{Density of valence electrons for the ground state ($7s^27p^2 \  (1/2,1/2)_0$) of Fl (E114).  
See Fig.~\ref{f:Uut} for notations.}
\label{f:Fl}
\end{figure}

\begin{figure}
\resizebox{0.42\textwidth}{!}{
\includegraphics{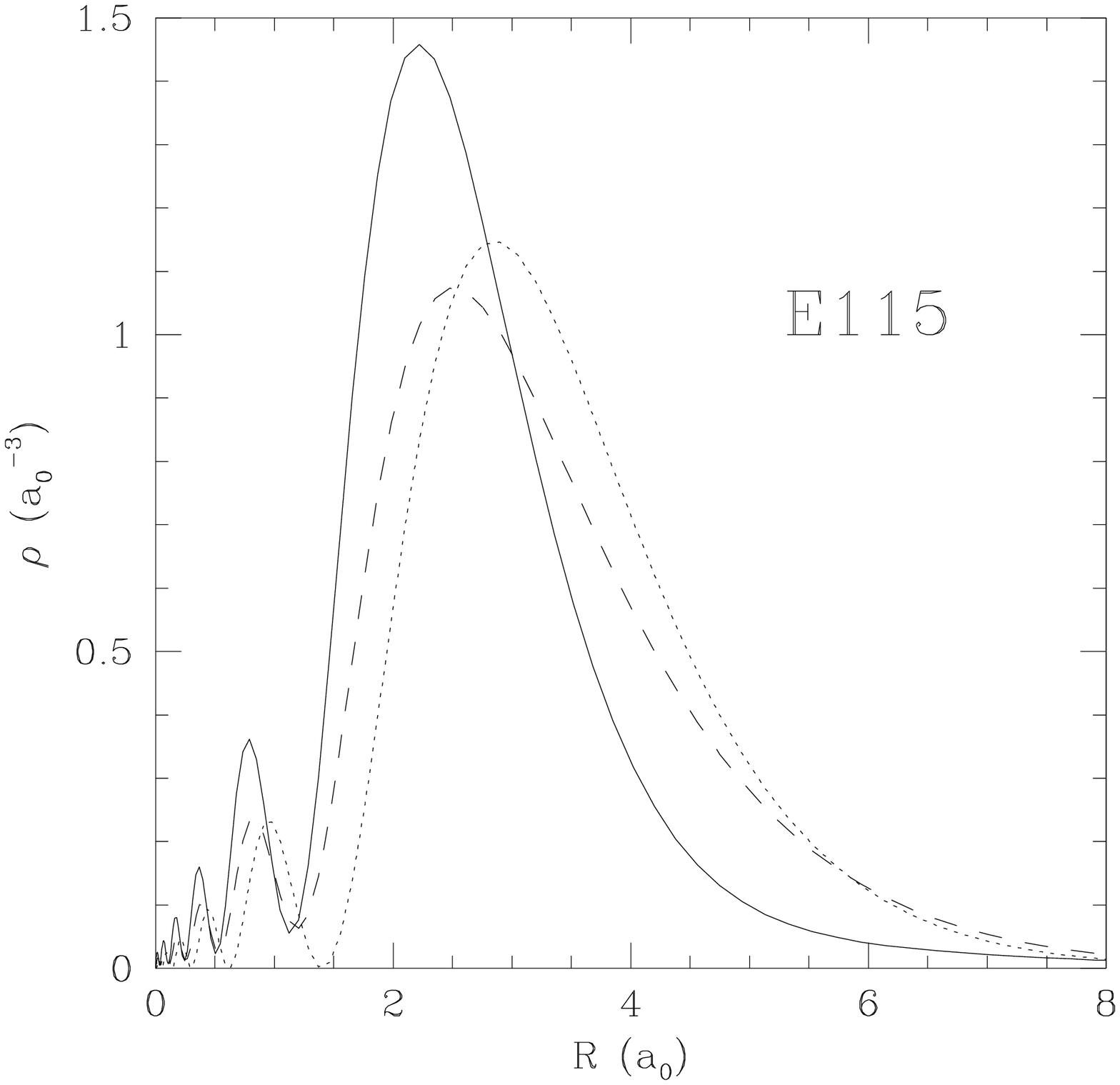}}
\caption{Density of valence electrons for the ground state ($7p^3 \ ^4$S$^o_{3/2}$) of Uup (E115). S 
See Fig.~\ref{f:Uut} for notations.}
\label{f:Uup}
\end{figure}

Calculated energy levels, $g$-factors and ionization potentials of superheavy elements 
E113, E114 and E115 and their ions are presented in Tables \ref{t:E113}, \ref{t:E114}, \ref{t:E115}. 
The results of earlier calculations are also presented for comparison.  As it was discussed 
above we expect that the accuracy of our calculated energies is about 1\%. In most
of cases the difference with other results is slightly larger, varying from about 1\% to few per
cents. Since other calculations must also have uncertainty, the results are rather in agreement
with each other. Note, that the agreement for E114 is significantly better than for E113 (see
Tables \ref{t:E113} and \ref{t:E114}).

The $g$-factors are useful for identification of states. We can see from the data in the tables that the 
values of calculated $g$-factors are close to corresponding non-relativistic values given by
\begin{equation}
g_{\rm NR} = 1 + \frac{J(J+1)-L(L+1)+S(S+1)}{2J(J+1)}.
\label{e:gf}
\end{equation}
This allows us to use non-relativistic labelling of the states.

Calculated IPs of the superheavy elements are compared with IPs of their lighter
analogs and with other calculations in Table~\ref{t:IP}. The agreement of present results
with experiment and with best earlier calculations is on the level of 1\%.

Comparing the spectra of superheavy elements and their lighter analogs (see tables 
\ref{t:E113}, \ref{t:E114}, \ref{t:E115} and \ref{t:Hg-Bi}) reveals the difference which can be
attributed to relativistic relaxation of the $s$ and $p_{1/2}$ states and increased fine structure intervals
for $p$ and $d$ states. For example, ionization potentials of E113 and E114 are significantly
larger than those of Tl and Pb. This is because IPs of these elements are the energies of removal 
of the external $p_{1/2}$ electron from the atom. However, due to relativistic relaxation, 
the $7p_{1/2}$ states of E113 and E114 are deeper on the energy scale than the $6p_{1/2}$ states
of Tl and Pb. In contrast, the IP of E115 is smaller than those of Bi. These is because these atoms 
have three external $p$-electrons which means that the $p_{3/2}$ state is also occupied. The
IP is the energy to remove the $p_{3/2}$ electron. Due to larger fine structure, the $7p_{3/2}$
electron of E115 is higher on the energy scale than the $6p_{3/2}$ electron of Bi.
Note that the E115$^+$ ion has no $7p_{3/2}$ electron and its ionization potential is larger than
for the case of E114$^+$ due to larger $Z$ (see Table \ref{t:IP}).

Similar tendency can be observed for polarizabilities (see Tables \ref{t:SHEpol} and \ref{t:pol}).
The polarizabilities of E113 and E114 are smaller than those of Tl and Pb, while polarizability 
of E115 is lager than those of Bi.

Figs. \ref{f:Uut}, \ref{f:Fl} and \ref{f:Uup} show electron density of valence electrons of superheavy elements
E113, E114 and E115 calculated in different approximations.  Solid line represents the result of most
accurate calculations using the CI wave function and formula (\ref{e:rho}). Dashed line shows the density
in the relativistic Hartree-Fock approximation calculated using formula (\ref{e:rhoHF}). Dotted line is the 
Hartree-Fock
density calculated in the non-relativistic limit. The difference between first two lines is due to correlations while
the difference between latter two lines is due to relativistic effects associated with Dirac equation (\ref{e:Dirac}).
Comparing the graphs for E113 and E114 shows that the relativistic effects cause significantly more change
to the electron density than correlations. Both atoms have the  $7s_{1/2}$  and $7p_{1/2}$ valence states 
with two electrons in the  $7s_{1/2}$ state and one or two electrons in the  $7p_{1/2}$ state.
Relativistic effects cause this states to contract towards the core. Configuration interaction brings 
admixture of other states most of which are also $p_{1/2}$ or $s_{1/2}$ states. They experience 
similar contraction causing little change in electron density.
The picture is more complicated for the E115 atom (Fig.~\ref{f:Uup}) due to the presence of the $7p_{3/2}$ state. 
Relativistic effects move the $7p_{1/2}$ and  $7p_{3/2}$ states in opposite directions, therefore, 
the change in total density is smaller while the effect of correlations is larger.  

\section{Conclusion}

Accurate calculations of energy levels, ionization potentials, polarizabilities and densities of valence electrons 
of superheavy elements E113, E114 and E115 are presented.  Similar calculations for lighter analogs of the
superheavy elements are presented for comparison and for the control of accuracy. The accuracy for the energies
is expected to be on the level of 1\% and the accuracy for the polarizabilities is few per cent.
The study of relativistic and correlation effects reveal that the difference in electron structure of superheavy
elements compared to their lighter analogs mostly come from relativistic effects associated with Dirac
equation. Inclusion of Breit and QED effects is important for accurate results but cause little difference.
The contribution of the correlations is similar for superheavy elements and their lighter analogs. 
Accurate treatment of the correlations represent the main challenge for the calculations.
For atoms with relatively simple electron structure (no more than three valence electrons),
as those considered in present paper, the calculations are done with sufficiently high accuracy. 

\begin{acknowledgement}
The work was supported in part by the Australian Research Council.
\end{acknowledgement}

%\bibliographystyle{apsrev}
%\bibliography{dzuba,super,other,pol}
\end{document}